\documentclass{article}

\usepackage{iclr2026_conference,times}

\usepackage{placeins}

\usepackage{latexsym}
\usepackage{amssymb}
\usepackage{amsmath}
\usepackage{bm}
\usepackage{microtype}
\usepackage{inconsolata}
\usepackage{float}     
\usepackage{needspace} 
\usepackage{lineno}

\usepackage{graphicx}
\usepackage{booktabs}
\usepackage{multirow}
\usepackage{makecell}
\usepackage{siunitx}
\usepackage[table]{xcolor}

\usepackage{enumitem}

\usepackage{hyperref}
\usepackage{url}

\usepackage{caption}

\usepackage[normalem]{ulem}
\usepackage{subcaption}

\newcommand{\revise}[1]{#1}

\begin{document}
\iclrfinalcopy

\title{LLM as a Risk Manager: LLM Semantic Filtering for Lead–Lag Trading in Prediction Markets}

\author
{\textbf{Sumin Kim}$^{1,*}$,
    \textbf{Minjae Kim}$^{1,*}$,
    \textbf{Jihoon Kwon}$^{1,*}$,
    \textbf{Yoon Kim}$^{2}$,
    \textbf{Nicole Kagan}$^{3}$,\\
    \textbf{Joo Won Lee}$^{4}$,
    \textbf{Oscar Levy}$^{5}$,
    \textbf{Alejandro Lopez-Lira}$^{6}$,
    \textbf{Yongjae Lee}$^{7,**}$,
    \textbf{Chanyeol Choi}$^{1,**}$\\
    \\$^{1}$LinqAlpha \quad
    $^{2}$Massachusetts Institute of Technology \quad
    $^{3}$Kalshi \quad \\
    $^{4}$Arrowpoint Investment Partners \quad
    $^{5}$University of California, Berkeley\quad \\
    $^{6}$University of Florida\quad
    $^{7}$UNIST \\
    [6pt]
    \footnotesize
    $^{*}$Equal Contribution. \quad $^{**}$Corresponding authors.
}
\maketitle

\begin{abstract}
Prediction markets provide a unique setting where event-level time series are directly tied to natural-language descriptions, yet discovering robust lead–lag relationships remains challenging due to spurious statistical correlations.
We propose a hybrid two-stage causal screener to address this challenge: (i) a statistical stage that uses Granger causality to identify candidate leader–follower pairs from market-implied probability time series, and (ii) an LLM-based semantic stage that re-ranks these candidates by assessing whether the proposed direction admits a plausible economic transmission mechanism based on event descriptions.
Because causal ground truth is unobserved, we evaluate the ranked pairs using a fixed, signal-triggered trading protocol that maps relationship quality into realized profit and loss (PnL).
On Kalshi Economics markets, our hybrid approach consistently outperforms the statistical baseline.
Across rolling evaluations, the win rate increases from 51.4\% to 54.5\%.
Crucially, the average magnitude of losing trades decreases substantially from $\$$649 to $\$$347.
This reduction is driven by the LLM's ability to filter out statistically fragile links that are prone to large losses, rather than relying on rare gains.
These improvements remain stable across different trading configurations, indicating that the gains are not driven by specific parameter choices.
Overall, the results suggest that LLMs function as \emph{semantic risk managers} on top of statistical discovery, prioritizing lead--lag relationships that generalize under changing market conditions.
\end{abstract}

\section{Introduction}
\label{sec:intro}

Real-world events rarely occur in isolation; information about one event
often changes expectations about what happens next.
For example, an inflation surprise can raise expectations of tighter
monetary policy.
These dependencies often manifest as \emph{lead--lag} structure, where
movement in one event systematically precedes movement in another
\citep{bennett2022leadlag}.

Identifying such structure from observational time series remains
challenging in practice.
Traditional statistical methods such as Granger causality can detect
directional predictability, but the resulting relationships are often
unstable over time and may not reflect meaningful underlying mechanisms
\citep{rossi2019gcinstability,spuriousGC2011}.
In large-scale settings, statistically significant lead--lag links
frequently fail to generalize out of sample and can induce substantial
losses when used in downstream decision-making.

Meanwhile, large language models (LLMs) have shown strong performance in
time-series forecasting tasks, including predicting future trajectories
from noisy observations \citep{gruver2023llmtime,jin2024timellm}.
However, these approaches typically focus on individual time series and
do not address whether relationships \emph{between} events are
meaningful or robust.
This raises a distinct question:
\emph{can LLMs help distinguish mechanistically plausible inter-event
relationships from brittle statistical correlations?}

Our key idea is to use LLMs as
\emph{semantic filters} that evaluate the plausibility of statistically
identified lead--lag relationships.
We study this question in a prediction-market setting, where event-level
time series continue to update over time and relationships can be
evaluated on future, unseen periods under changing market conditions.

We implement a two-stage framework.
First, we apply Granger causality to identify candidate
leader--follower event pairs from market-implied probability time series(Statistical Approach).
Second, we use an LLM to re-rank these candidates based solely on the
events’ natural-language descriptions, prioritizing relationships that
admit a plausible real-world mechanism(Hybrid Approach).

Across 18 rolling evaluations on Kalshi Economics markets, LLM-based
semantic filtering consistently improves trading performance relative to
Granger screening alone.
At a 7-day holding horizon, total PnL more than doubles, driven mainly by a pronounced reduction in downside
risk.
These results suggest that LLMs act as robustness filters on top of
statistical discovery, helping to exclude fragile lead--lag relationships
that fail under changing market conditions.

\section{Related Work}
\label{sec:related}

\paragraph{Limits of statistical significance in lead--lag discovery.}
Lead--lag patterns, in which movements in one time series appear to precede those in another, are widely documented in economics and finance \citep{lo1990nonsynchronous,hasbrouck1995information}.
Motivated by these empirical regularities, a large body of work seeks to formalize directional relationships using statistical discovery tools such as Granger causality and related predictive tests \citep{granger1969investigating,exactRecoveryPairwiseGC2023,chaudhry2017fdrgc}.

However, even when statistically significant, many discovered lead--lag relationships remain fundamentally ambiguous and limited.
In large-scale settings, pairwise screening inevitably produces links that survive selection purely by chance, a concern long emphasized in the data-snooping and multiple-testing literature \citep{sullivan1999datasnooping}.
These issues are further exacerbated when relationships evolve over time or exhibit structural breaks, leading to instability in Granger-causal links across samples \citep{psaradakis1995gcstructuralbreaks,rossi2019gcinstability,spuriousGC2011}.
As a result, statistical significance alone provides limited guidance for distinguishing mechanistically meaningful inter-event relationships from brittle correlations.

To assess robustness, discovered relationships are therefore often evaluated using out-of-sample predictive performance or rolling evaluation schemes \citep{diebold1995accuracy,rossi2021jel_instabilities}.
Yet even when validated out of sample, such evaluations provide limited guidance about whether an identified lead--lag direction corresponds to a coherent real-world transmission mechanism.

\paragraph{Semantic evaluation with LLMs and event-based settings.}
Recent work has explored the use of large language models (LLMs) beyond traditional language tasks, including evaluating explanations, relationships, and counterfactuals expressed in natural language \citep{zheng2023llmasjudge,yang2024causalbenchreview,causalbench2024}.
In these settings, LLMs are not used to infer causal structure from data, but rather to assess the coherence or plausibility of candidate relationships.
This perspective is particularly relevant for lead--lag analysis, where temporal ordering alone is insufficient to establish whether a relationship admits a meaningful real-world transmission mechanism.

Prediction markets provide a natural platform for this type of evaluation.
Each contract corresponds to a clearly defined event described in natural language and yields a time series of market-implied probabilities reflecting collective expectations \citep{wolfers2004predictionmarkets,berg2008predictionaccuracy}.
This structure enables the study of inter-event relationships while allowing discovered dependencies to be evaluated on future, unseen periods under changing market conditions.


\section{Background}
\label{sec:background}

This section introduces the key concepts and notation underlying our
analysis of directional relationships between event-level time series.
Our objective is to study directional lead–lag relationships between event-level time series and establish a framework for evaluating their predictive relevance.

We formally define directional lead--lag
relationships in Section.~\ref{subsec:bg_leadlag}.
Section~\ref{subsec:bg_predictionmarket} describes pre-processing procedure of time-series data, and Section~\ref{subsec:bg_granger} introduces Granger causality as a statistical tool for operationalizing directional predictability.
Finally, Section~\ref{subsec:bg_sign} defines how to annotate discovered lead--lag relationships.

\subsection{Directional lead--lag relationships}
\label{subsec:bg_leadlag}

Let $\{x_t\}_{t=1}^T$ and $\{y_t\}_{t=1}^T$ be two real-valued time series
indexed by time $t$.
We say that $x$ \emph{leads} $y$ if past values of $x_t$ contain predictive
information about future values of $y_t$ beyond what is contained in the
past of $y_t$ itself.

Formally, $x$ leads $y$ if there exists a forecast horizon $h>0$ such that
\[
\mathbb{E}\!\left[y_{t+h} \mid x_{1:t}, y_{1:t}\right]
\neq
\mathbb{E}\!\left[y_{t+h} \mid y_{1:t}\right].
\]
This definition is directional and asymmetric: $x$ leading $y$ does not
imply that $y$ leads $x$. Throughout this paper, we study ordered event pairs exhibiting directional lead--lag structure—also referred to as leader--follower pairs $(L,F)$.

\subsection{Time-series representation of prediction-market prices}
\label{subsec:bg_predictionmarket}

Prediction markets trade contracts whose payoff depends on the
realization of a future event.
For each event $i$, the daily YES price forms an event-level time series
that reflects the market’s collective belief about the probability of
that event occurring.

Let $p_{i,t} \in [0,100]$ denote the daily YES price (in percentage
points) of event $i$ at time $t$.
To obtain an unbounded real-valued signal suitable for time-series
analysis, we apply the log-odds transformation
\begin{equation}
\label{eq:logodds}
\ell_{i,t}
=
\log\!\left(\frac{p_{i,t}}{100 - p_{i,t}}\right),
\end{equation}
where $\ell_{i,t} \in \mathbb{R}$.

This transformation mitigates boundary effects near $0$ and $100$ and
renders price changes approximately additive, which facilitates
statistical modeling.

\subsection{Directional predictability via Granger causality}
\label{subsec:bg_granger}

To test lead--lag relationships, we require a statistical notion of
directional predictability.
We use Granger causality, which assesses whether incorporating the
lagged history of one series improves prediction of another beyond what
is explained by the target series’ own past.

Let $\{x_t\}_{t=1}^T$ and $\{y_t\}_{t=1}^T$ be two time series, and let
$p$ denote the chosen lag length in the autoregressive model.
Granger causality tests are typically formulated under the assumption
that the input series are approximately stationary.
In practice, stationarity is commonly assessed using unit-root tests
such as the Augmented Dickey--Fuller (ADF) test, which evaluates whether
a time series exhibits a unit root (i.e., behaves like a random walk).
When non-stationarity is detected, differencing is a common remedy
\citep{dickeyfuller1979,saiddickey1984,hamilton1994}.

\revise{A standard approach to testing Granger causality is to use vector
autoregressive (VAR) models, which capture linear dependencies between
multiple time series through their lagged values \citep{hamilton1994}.
In the bivariate case, $y_t$ can be modeled with optional lags of $x_t$
as}

\begin{equation}
\label{eq:granger_var}
y_t
=
\alpha_0
+ \sum_{k=1}^{p}\alpha_k\,y_{t-k}
+ \sum_{k=1}^{p}\beta_k\,x_{t-k}
+ \varepsilon_t,
\end{equation}
where $\varepsilon_t$ is a mean-zero error term.

The Granger test evaluates the null hypothesis
$H_0:\beta_1=\cdots=\beta_p=0$, corresponding to the absence of directional predictability from $x$ to $y$ under lag length $p$. 
Rejection of $H_0$ indicates a lead--lag relationship from $x$ to $y$.
Thus, Granger causality provides a testable proxy for the theoretical lead--lag definition in Section~\ref{subsec:bg_leadlag}.

\subsection{Sign of co-movement}
\label{subsec:bg_sign}

Beyond temporal direction, lead--lag relationships can be further
characterized by whether the two series tend to move in the same or
opposite directions.
For an ordered lead--lag (leader--follower) pair $(L \rightarrow F)$ with corresponding time series $(x_{L,t}, x_{F,t})$, we define the sign of co-movement as
\begin{equation}
\label{eq:sign}
s
=
\mathrm{sgn}\!\left(\mathrm{corr}(x_{L,t}, x_{F,t})\right)
\in \{-1, +1\},
\end{equation}
where $\mathrm{corr}(\cdot,\cdot)$ denotes the Pearson correlation
coefficient. Here, $s=+1$ indicates aligned movement and $s=-1$ indicates opposing
movement, conditional on the identified lead--lag ordering.
This quantity is purely descriptive and does not imply any causal or structural interpretation.
It indicates whether increases in the leading series tend to be associated
with increases or decreases in the lagging series.

\section{Methods}
\label{sec:method}

\revise{
Recall that our study aims to assess whether LLMs can distinguish causal relationships that are mechanistically meaningful.
To assess this question, we compare two methods for ranking candidate causal relationships: a purely \emph{statistical} approach, and a \emph{hybrid} approach that applies LLM-based re-ranking on top of the statistical approach.
We evaluate each ranking using the same trading protocol, where investment performance reflects the reliability of the identified causal relationships.
}

\revise{
In this section, we introduce the two ranking methods: \emph{statistical} and \emph{hybrid} (Section.~\ref{sec:ranking_methods}), and describe the trading protocol (Section.~\ref{sec:trading_protocol}).
}
\subsection{Ranking Methods for Candidate Causal Relationships}
\label{sec:ranking_methods}

\begin{figure*}[t]
  \centering
  \includegraphics[width=0.9\textwidth]{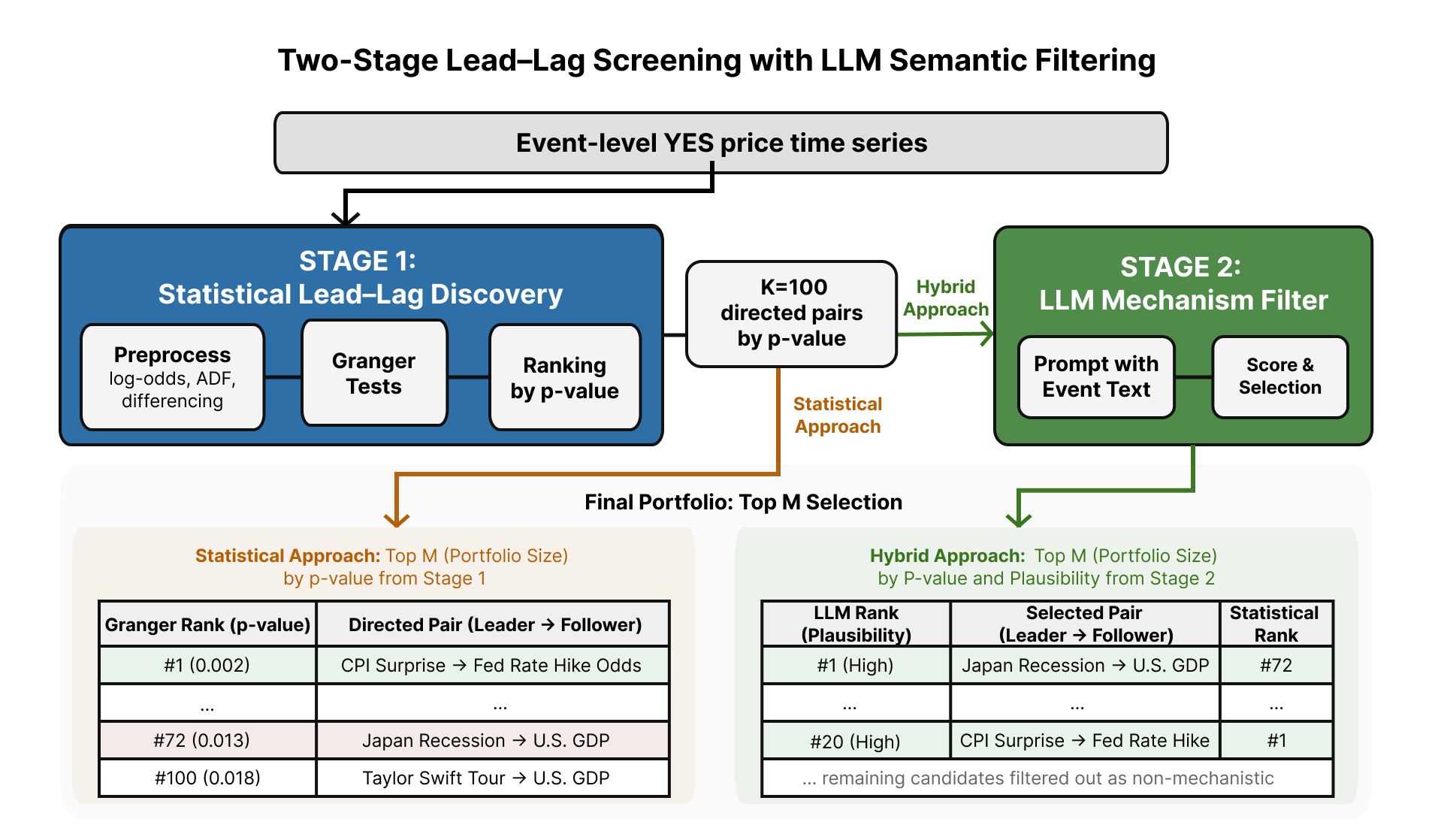}
\caption{
Two-stage framework for leader–follower pair discovery in prediction markets. Stage 1 produces a candidate set of Top K directed pairs (K=100) ranked by Granger significance, and Stage 2 applies LLM-based semantic re-ranking to select the final Top M portfolio (M=20).
}
  \label{fig:framework}
\end{figure*}

Figure~\ref{fig:framework} summarizes the two-stage causal filtering framework used in this study.
Starting from event-level prediction market price time series, the pipeline consists of
(i) a statistical filtering stage that identifies candidate leader--follower relationships based on time-series evidence,
and (ii) a semantic filtering stage that re-ranks these candidates using an LLM.
The two stages correspond to the \emph{Statistical Approach} and \emph{Hybrid Approach} described below, and both produce ranked lists of directed event pairs that are evaluated using the same trading protocol.

For each event $i$, we obtain a daily price time series
$\mathcal{P}_i = \{p_{i,t}\}_{t=1}^T$, where $p_{i,t}$ denotes the market-implied price at time $t$ and $T$ is the length of the observation window.
From $\mathcal{P}_i$, we construct a stationary market signal
$\mathcal{X}_i = \{x_{i,t}\}_{t=1}^T$ by transforming prices into log-odds. \revise{Standard stationarity preprocessing is applied prior to Granger testing.}

\paragraph{Statistical Approach.}
Given a pair of events $(i,j)$, the statistical approach aims to determine whether there exists a directional lead--lag relationship between the two events based \emph{purely on time-series evidence}.
To this end, we apply Granger causality tests to the stationary market series constructed from their daily price data.

For each unordered event pair $\{i,j\}$, we evaluate both possible causal
directions, $(i \rightarrow j)$ and $(j \rightarrow i)$.
In each direction, we estimate vector autoregressive (VAR) models over multiple lag lengths and assess whether the lagged history of the candidate leader provides incremental predictive power for the follower beyond its own past.

We retain the direction exhibiting stronger statistical evidence of
directional predictability as the candidate causal relationship for the
event pair.
Finally, all candidate relationships are ranked by their statistical
strength, yielding an ordered list of directed leader--follower pairs
from strongest to weakest evidence of Granger-based predictability.

\paragraph{Hybrid Approach.}
The hybrid approach takes as its starting point the set of candidate causal relationships identified by the statistical approach. 
The role of the LLM is not to discover new relationships, but to re-rank these statistically validated candidates based on semantic and mechanism-level plausibility.
In doing so, the approach preserves statistical validity while adjusting priorities to favor relationships that have coherent economic interpretations.

Specifically, for each directed event pair retained by the statistical approach, we prompt a LLM with the event titles and descriptions and ask whether a plausible economic mechanism exists by which movement in the leader could precede movement in the follower.
The model evaluates the semantic coherence of the proposed causal direction and assigns a plausibility score, with higher scores indicating stronger mechanism-level support.

We then use these plausibility scores to re-rank the statistically screened candidate relationships.
The final hybrid ranking thus preserves the statistical validity ensured by Granger causality while prioritizing relationships that are also supported by coherent economic reasoning.

\subsection{Trading-Based Evaluation of Ranking Methods}
\label{sec:trading_protocol}

\begin{figure*}[t]
  \centering
  \includegraphics[width=0.9\textwidth]{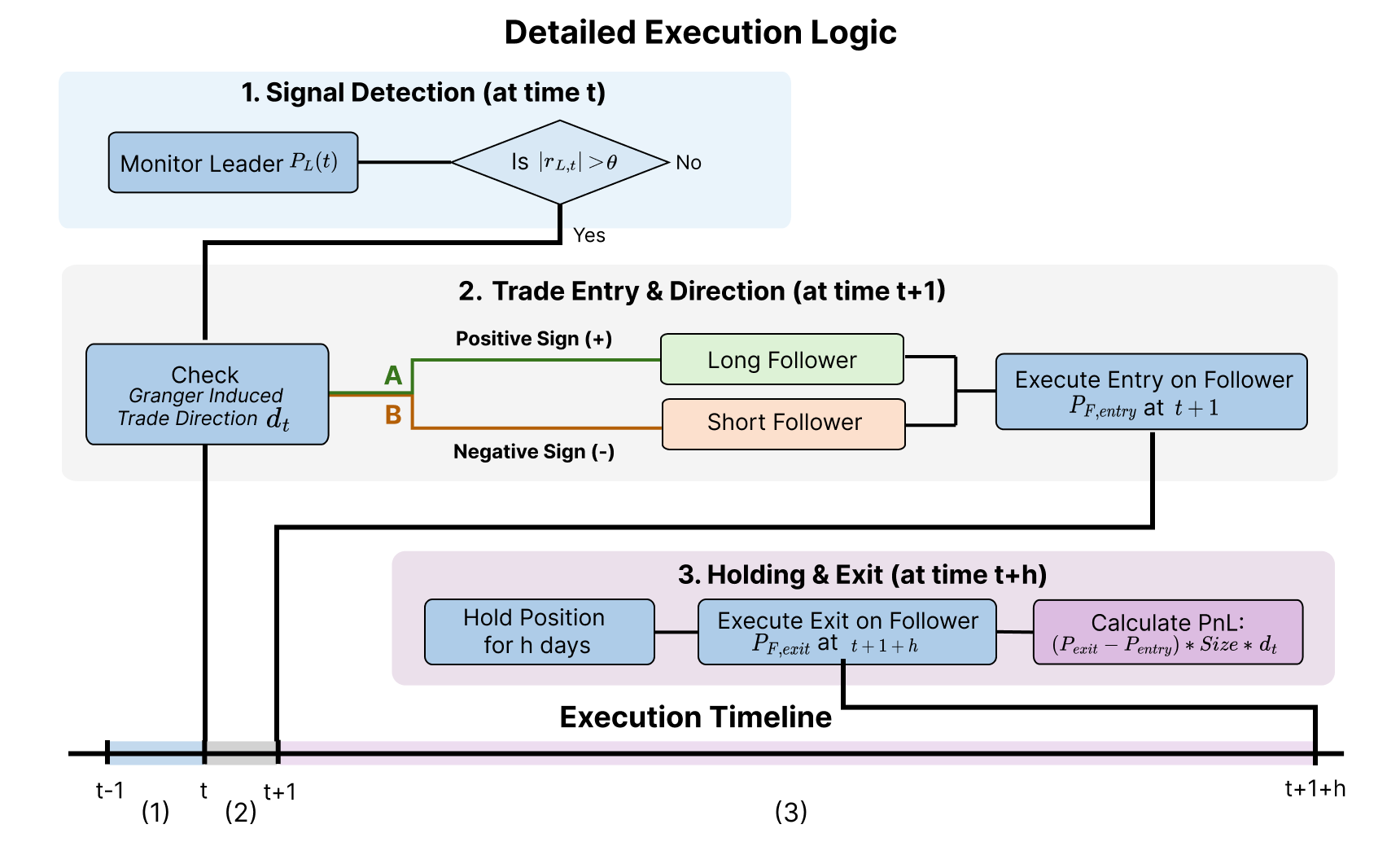}
  \caption{
Signal-triggered trading protocol used to evaluate ranked lead-lag relationships from Figure~\ref{fig:framework}: leader price moves trigger follower trades, with direction determined by the Granger-induced trade sign and out-of-sample PnL used to evaluate the ranked pair list.
}
  \label{fig:trade}
\end{figure*}
Since the ground truth of causal relationships is not directly observable, evaluating the quality of different relationship rankings is inherently challenging. To address this issue, we evaluate the ranking approaches based on trading performance.
The trading rule is intentionally designed to be simple and signal-driven: when a sufficiently large price movement occurs in a leader event, the strategy mechanically follows the corresponding follower.
Therefore, differences in realized trading performance can be attributed to differences in the quality of the underlying causal rankings. Figure~\ref{fig:trade} illustrates the resulting signal-based trade lifecycle, which we detail step by step below.

\paragraph{Step 1: leader-based trigger}
Consider a directed event pair $(L,F)$, where $L$ denotes the leader event and $F$ the corresponding follower as identified by a causal ranking.
At each time $t$, we monitor the daily price movement of the leader event and trigger a trading signal when the magnitude of the price change exceeds a predefined threshold. \revise{Formally, let $p_{L,t}$ denote the market-implied price (in percent) of the leader event at time $t$. We define the one-day relative price change as
$r_{L,t} \triangleq (p_{L,t}-p_{L,t-1})/p_{L,t-1}$.
A trading signal is generated if $|r_{L,t}| > \theta$,
where $\theta$ is a fixed threshold in decimal form.}

\paragraph{Step 2: Follower Entry.}
When a trading signal is generated for a leader--follower pair $(L,F)$ at time $t$, we enter a position in the follower event at time $t+1$.
We define the \emph{Granger-induced trade direction} as
\[
d_t = \mathrm{sign}(r_{L,t}) \times s(L \rightarrow F),
\]
where $s(L \rightarrow F)$ is the sign of the identified Granger relationship.
If $d_t = +1$, we take exposure to an increase in the follower event probability (buy YES shares), and if $d_t = -1$, we take exposure to a decrease in the follower event probability (sell YES shares).
This construction follows directly from the lead--lag interpretation of the Granger sign described in Section~\ref{subsec:bg_leadlag}.

\paragraph{Step 3: Holding Period and Exit.}
Once entered, the follower position is held for a fixed horizon of $h$ days and is exited mechanically at time $t+h+1$.
The realized profit or loss is computed based on the change in the follower's market-implied price over the holding period.

\section{Experiments}
\label{sec:experiments}

\subsection{Experimental Setup}
\label{subsec:exp_setup}

\paragraph{Data and windows.}
We use Kalshi prediction-market data in the Economics category spanning October 2021 to November 2025.
After filtering markets with insufficient activity or negligible price variation, we retain 554 event markets.
Evaluation follows a rolling-window protocol with 60-day training windows and 30-day testing windows, yielding 18 non-overlapping test periods. 

\paragraph{Pipeline and hyperparameters.}
\revise{Within each training window, we perform ADF unit-root testing and apply
first differencing when necessary prior to Granger analysis.
We then run pairwise Granger causality tests across all available markets.
For each event pair, we evaluate both causal directions and sweep lag
lengths $L \in \{1,2,3,4,5\}$, retaining the strongest statistical evidence
of directional predictability.
Because these optimal lag orders are short in market-implied probability dynamics, we treat the trading holding horizon $h$ as an independent evaluation parameter and report robustness across multiple horizons.}
From these tests, we retain the top $K=100$ directed pairs ranked by
statistical significance.
From this candidate set, we form a portfolio by selecting the top $M=20$
directed pairs under two ranking conditions: the \emph{statistical}
approach and the \emph{hybrid} approach (see Section~\ref{sec:ranking_methods}). \revise{For the semantic re-ranking stage, we use \texttt{GPT-5-nano} as the LLM. Each trade is executed with position size of 100 per contract. }

\paragraph{Evaluation settings and default configuration.}
We evaluate trading performance using holding horizons $h \in \{1,3,5,7,10,14,21\}$ days and report standard trading metrics including win rate, average win/loss per trade, and total PnL aggregated across test windows.
Unless otherwise specified, we report results under a default configuration consisting of a 7-day holding period, trades triggered when leader relative price change exceeds threshold $\theta = 0$
  (i.e., $|r_{L,t}| > 0$), and portfolios of size $M = 20$.

\subsection{Results}
\label{subsec:results}

\paragraph{LLM filtering yields substantial reductions in downside risk.}
Table~\ref{tab:main} reports the primary comparison between the statistical and hybrid approaches under the default evaluation setting.
LLM-based re-ranking leads to a substantial improvement in total PnL (+205\%) accompanied by a modest increase in win rate.
This pattern indicates that the gains are not driven by a small number of outlier wins or aggressive upside bets.
Instead, the dominant driver of the performance improvement is a large reduction in average loss magnitude (46.5\%),
highlighting enhanced downside control as the primary mechanism.

\paragraph{Loss reduction persists for distinct event pairs.}
To examine whether the observed loss reduction is driven by trivial semantic overlap, we analyze performance composition under the default setting.
As shown in Table~\ref{tab:decomp}, LLM filtering reduces average loss magnitude by 48.1\% for different-event pairs (from \$642 to \$333), comparable to the reduction observed for same-event pairs (42.9\%).
This indicates that the gains are not attributable to trivial event overlap, but persist for genuinely distinct event relationships.

\begin{table*}[t]
\centering

\caption{
Trading-based evaluation summary.
We compare portfolios constructed under the \emph{statistical approach} and the \emph{hybrid approach} across rolling test windows.
Metrics include trade count, win rate, average win/loss per trade, and total PnL; performance gains are driven primarily by reductions in average loss magnitude.
}
\label{tab:main}

\setlength{\tabcolsep}{6pt}
\begin{tabular}{lrrr}
\toprule
Metric & Statistical Approach & Hybrid Approach & Change \\
\midrule
Win Rate & 51.4\% & 54.5\% & +3.1pp \\
Avg Win & \$724 & \$636 & -12\% \\
Avg Loss & -\$649 & -\$347 & +46.5\% \\
\textbf{Total PnL} & \textbf{\$4,100} & \textbf{\$12,500} & \textbf{+205\%} \\
\bottomrule
\end{tabular}
 
\vspace{10pt}

\caption{
Comparison of downside losses across approaches under same-event and different-event settings.
Same-event pairs correspond to the same underlying real-world event with different thresholds or contract formulations, whereas different-event pairs correspond to genuinely distinct events.
We report average loss per trade; lower average loss indicates better downside control.
}
\label{tab:decomp}

\setlength{\tabcolsep}{8pt}
\begin{tabular}{lcc}
\toprule
Method & Same-Event & Different-Event \\
\midrule
Avg. Loss of Statistical Approach & -\$700 & -\$642 \\
Avg. Loss of Hybrid Approach     & -\$400 & -\$333 \\
\midrule
Loss Reduction       & 42.9\% & 48.1\% \\
\bottomrule
\end{tabular}
\vspace{10pt}


\caption{
Win rate (WR) by the magnitude of leader movement, indicating which side outperforms in investment outcomes.
Here, the magnitude of leader movement is defined as the amount of change in the leader event’s price $p$ at a given time.
Win rates (WR) are reported for Statistical Apporach (Statistical WR) and Hybrid Approach(Hybrid WR) portfolios; $\Delta$ denotes the win-rate difference in percentage points.
}
\label{tab:movement}

\setlength{\tabcolsep}{4pt}
\resizebox{0.8\textwidth}{!}{
\begin{tabular}{crrrc}
\toprule
Magnitude of leader movement (pt) & Statistical WR & Hybrid WR & Winner & $\Delta$ \\
\midrule

\hspace{1em}5--10 pt 
& 57.1\% & 66.7\% & LLM & \textbf{+9.5pp} \\
\hspace{1em}10+ pt 
& 53.8\% & 71.4\% & LLM & \textbf{+17.6pp} \\
\bottomrule
\end{tabular}}

\end{table*}

\paragraph{Semantic filtering is most valuable during large leader moves.}
We next examine performance by the magnitude of leader price changes.
As shown in Table~\ref{tab:movement}, semantic re-ranking delivers larger gains when leader markets experience larger discrete repricings.
For 5--10pt moves, win rate increases from 57.1\% to 66.7\% (+9.5pp),
and for moves exceeding 10pt, it rises from 53.8\% to 71.4\% (+17.6pp).
This pattern suggests that LLM filtering is particularly effective at mitigating regime-sensitive failures that arise during large market moves.

\paragraph{LLM rescues high-value relationships missed by Granger ranking.}
Beyond aggregate performance gains, the semantic filtering stage surfaces economically coherent lead--lag relationships that would be excluded under purely statistical or surface-level semantic screening.
Table~\ref{tab:representative_pairs} summarizes representative \emph{LLM-only} trades that fall outside the Granger Top-$M$ portfolio by $p$-value ranking, yet yield positive realized outcomes after LLM-based filtering.
Notably, the Japan Recession $\rightarrow$ U.S. GDP pair has a Granger rank of \#71—far outside the Top-$M$ cutoff—yet is selected by the LLM based on a clear cross-border macroeconomic mechanism.
Despite exhibiting low surface-form similarity and weak Granger statistics, the LLM assigns a negative causal sign, reasoning that
\emph{“recessions imply weaker domestic demand, and through cross-border trade, financial linkages, and policy spillovers, downturns in major economies tend to drag on aggregate growth.”}
Rather than relying on lexical or topical overlap, the LLM promotes such pairs by inferring latent economic transmission channels and assigning a consistent causal sign that translates into profitable trade directions.
These qualitative results illustrate that the LLM stage does not merely refine statistical rankings, but actively recovers structurally meaningful relationships that are brittle under both Granger-based ordering and lightweight embedding similarity.
\begin{table}[t]
  \centering
  \caption{
  Representative pairs ranked low by the Statistical Approach but elevated by the Hybrid Approach.
  SR = Statistical Rank (Granger $p$-value); HR = Hybrid Rank (after LLM re-ranking).
  }
  \small
  \setlength{\tabcolsep}{5pt}
  \begin{tabular}{llccc}
  \toprule
  Leader & Follower & SR & HR & PnL \\
  \midrule
  China 2022 GDP Growth $>$5\%     & World 2022 GDP Growth $>$3\%     & \#23 & \#17 & +\$1{,}100 \\
  Japan 2026 Recession           & US Q1 2025 GDP Growth $>$2\%     & \#71 & \#5  & +\$700 \\
  US 2025 Oil Prod. $>$14.5M bbl/day   & Brazil 2025 Inflation $>$5.5\%   & \#24 & \#15 & +\$600 \\
  India 2026 Recession           & CRE Delinq. Q4 2024 $>$3\%       & \#51 & \#3  & +\$200 \\
  \bottomrule
  \end{tabular}
  \label{tab:representative_pairs}
  \end{table}

\begin{table*}[t]
  \caption{
  Hold period ablation (trades with $|\Delta p_{L,t}|>0$ only).
  We report win rate and average loss for Statistical Approach versus Hybrid Approach portfolios as the holding horizon varies;
  Loss Reduction is the relative decrease in average loss magnitude under Hybrid filtering.
  }
    \label{tab:ablation}
  \centering
  \small
  \setlength{\tabcolsep}{8pt}
  \begin{tabular}{lrrrrc}
  \toprule
  Hold & Stat. WR & Hybrid WR & Stat. Avg Loss & Hybrid Avg Loss & Loss Reduction \\
  \midrule
  1d  & 56.0\% & 66.7\% & -\$536 & -\$283 & 47.2\% \\
  3d  & 44.7\% & 62.8\% & -\$773 & -\$469 & 39.4\% \\
  5d  & 47.5\% & 66.0\% & -\$745 & -\$500 & 32.9\% \\
  \textbf{7d}  & \textbf{51.4\%} & \textbf{54.5\%} & \textbf{-\$649} & \textbf{-\$347} & \textbf{46.5\%} \\
  10d & 51.4\% & 56.9\% & -\$497 & -\$371 & 25.3\% \\
  14d & 49.5\% & 56.1\% & -\$661 & -\$400 & 39.5\% \\
  21d & 47.3\% & 53.1\% & -\$973 & -\$753 & 22.6\% \\
  \midrule
  Mean & $\sim$50\% & $\sim$59\% & -\$689 & -\$446 & 36.2\% \\
  \bottomrule
  \end{tabular}
  \end{table*}

\subsection{Robustness and Ablations}
\label{subsec:robustness}



We assess the robustness of this loss-reduction behavior by varying the holding horizon while keeping all other evaluation settings fixed.
Table~\ref{tab:ablation} reports performance across holding periods ranging from 1 to 21 days.
Across all tested horizons, LLM filtering consistently reduces average loss magnitude relative to the statistical baseline.
The relative reduction in average loss ranges from 22.6\% to 47.2\%, with a mean reduction of 36.2\% across holding periods.
This consistency indicates that the observed behavior is not driven by a particular choice of holding horizon, but reflects a stable enhancement in the robustness of the selected relationships.
While win rates vary with the holding period, the loss-reduction effect persists even when win-rate gains are modest, further supporting the interpretation that semantic filtering primarily improves downside control.

\section{Conclusion}
We introduced a dual-stage causal filtering framework for prediction markets, combining statistical lead--lag screening via Granger causality with LLM-based semantic verification.
Across rolling-window evaluations, LLM re-ranking consistently improves trading performance, with gains driven primarily by substantial reductions in average loss magnitude (downside risk) and accompanied by modest improvements in win rate. Taken together, these results suggest that LLMs can act as semantic risk managers on top of statistical discovery, prioritizing relationships that generalize more reliably under changing market conditions.

\bibliography{main}

@article{bennett2022leadlag,
  title   = {Lead--lag detection and network clustering for multivariate time series with an application to the US equity market},
  author  = {Bennett, Stefanos and Cucuringu, Mihai and Reinert, Gesine},
  journal = {Machine Learning},
  volume  = {111},
  pages   = {4497--4538},
  year    = {2022},
  doi     = {10.1007/s10994-022-06250-4}
}

@article{berg2008predictionaccuracy,
  title={Prediction market accuracy in the long run},
  author={Berg, Joyce E and Nelson, Forrest D and Rietz, Thomas A},
  journal={International Journal of Forecasting},
  volume={24},
  number={2},
  pages={285--300},
  year={2008},
  publisher={Elsevier}
}

@inproceedings{chaudhry2017fdrgc,
  title={Uncertainty assessment and false discovery rate control in high-dimensional Granger causal inference},
  author={Chaudhry, Aditya and Xu, Pan and Gu, Quanquan},
  booktitle={International Conference on Machine Learning},
  pages={684--693},
  year={2017},
  organization={PMLR}
}

@article{lo1990nonsynchronous,
  title={An econometric analysis of nonsynchronous trading},
  author={Lo, Andrew W and MacKinlay, A Craig},
  journal={Journal of Econometrics},
  volume={45},
  number={1-2},
  pages={181--211},
  year={1990},
  publisher={Elsevier}
}

@article{hasbrouck1995information,
  title={One security, many markets: Determining the contributions to price discovery},
  author={Hasbrouck, Joel},
  journal={The journal of Finance},
  volume={50},
  number={4},
  pages={1175--1199},
  year={1995},
  publisher={Wiley Online Library}
}

@article{wolfers2004predictionmarkets,
  title   = {Prediction Markets},
  author  = {Wolfers, Justin and Zitzewitz, Eric},
  journal = {Journal of Economic Perspectives},
  volume  = {18},
  number  = {2},
  pages   = {107--126},
  year    = {2004}
}

@article{granger1969investigating,
  title={Investigating causal relations by econometric models and cross-spectral methods},
  author={Granger, Clive WJ},
  journal={Econometrica: journal of the Econometric Society},
  pages={424--438},
  year={1969},
  publisher={JSTOR}
}

@article{psaradakis1995gcstructuralbreaks,
  title={Granger-causality in the presence of structural breaks},
  author={Ventosa-Santaul{\`a}ria, Daniel and Vera-Vald{\'e}s, J Eduardo},
  journal={Economics Bulletin},
  volume={3},
  number={61},
  year={2008},
  publisher={Economics Bulletin}
}

@article{rossi2019gcinstability,
  title   = {Vector autoregressive-based Granger causality tests in the presence of instabilities},
  author  = {Rossi, Barbara and Wang, Yu},
  journal = {The Stata Journal},
  volume  = {19},
  number  = {4},
  pages   = {883--899},
  year    = {2019},
  doi     = {10.1177/1536867X19893631}
}

@article{spuriousGC2011,
  title={Understanding spurious regressions in econometrics},
  author={Phillips, Peter CB},
  journal={Journal of econometrics},
  volume={33},
  number={3},
  pages={311--340},
  year={1986},
  publisher={Elsevier}
}

@article{sullivan1999datasnooping,
  title={Data-snooping, technical trading rule performance, and the bootstrap},
  author={Sullivan, Ryan and Timmermann, Allan and White, Halbert},
  journal={The journal of Finance},
  volume={54},
  number={5},
  pages={1647--1691},
  year={1999},
  publisher={Wiley Online Library}
}

@article{exactRecoveryPairwiseGC2023,
  title={Exact recovery of Granger causality graphs with unconditional pairwise tests},
  author={Kinnear, Ryan J and Mazumdar, Ravi R},
  journal={Network Science},
  volume={11},
  number={3},
  pages={431--457},
  year={2023},
  publisher={Cambridge University Press}
}

@article{diebold1995accuracy,
  title={Comparing predictive accuracy},
  author={Diebold, Francis X and Mariano, Robert S},
  journal={Journal of Business \& economic statistics},
  volume={20},
  number={1},
  pages={134--144},
  year={2002},
  publisher={Taylor \& Francis}
}

@article{rossi2021jel_instabilities,
  title={Forecasting in the presence of instabilities: How we know whether models predict well and how to improve them},
  author={Rossi, Barbara},
  journal={Journal of Economic Literature},
  volume={59},
  number={4},
  pages={1135--1190},
  year={2021},
  publisher={American Economic Association 2014 Broadway, Suite 305, Nashville, TN 37203-2425}
}

@article{dickeyfuller1979,
  title   = {Distribution of the Estimators for Autoregressive Time Series with a Unit Root},
  author  = {Dickey, David A. and Fuller, Wayne A.},
  journal = {Journal of the American Statistical Association},
  volume  = {74},
  number  = {366},
  pages   = {427--431},
  year    = {1979},
  doi     = {10.1080/01621459.1979.10482531}
}

@article{saiddickey1984,
  title={Testing for unit roots in autoregressive-moving average models of unknown order},
  author={Said, Said E and Dickey, David A},
  journal={Biometrika},
  volume={71},
  number={3},
  pages={599--607},
  year={1984},
  publisher={Oxford University Press}
}

@book{hamilton1994,
  title={Time series analysis},
  author={Hamilton, James D},
  year={2020},
  publisher={Princeton university press}
}

@article{gruver2023llmtime,
  title={Large language models are zero-shot time series forecasters},
  author={Gruver, Nate and Finzi, Marc and Qiu, Shikai and Wilson, Andrew G},
  journal={Advances in Neural Information Processing Systems},
  volume={36},
  pages={19622--19635},
  year={2023}
}

@inproceedings{jin2024timellm,
 author = {Jin, Ming and Wang, Shiyu and Ma, Lintao and Chu, Zhixuan and Zhang, James and Shi, Xiaoming and Chen, Pin-Yu and Liang, Yuxuan and Li, Yuan-Fang and Pan, Shirui and Wen, Qingsong},
 booktitle = {International Conference on Learning Representations},
 editor = {B. Kim and Y. Yue and S. Chaudhuri and K. Fragkiadaki and M. Khan and Y. Sun},
 pages = {23857--23880},
 title = {Time-LLM: Time Series Forecasting by Reprogramming Large Language Models},
 url = {https://proceedings.iclr.cc/paper_files/paper/2024/file/680b2a8135b9c71278a09cafb605869e-Paper-Conference.pdf},
 volume = {2024},
 year = {2024}
}

@article{zheng2023llmasjudge,
  title={Judging llm-as-a-judge with mt-bench and chatbot arena},
  author={Zheng, Lianmin and Chiang, Wei-Lin and Sheng, Ying and Zhuang, Siyuan and Wu, Zhanghao and Zhuang, Yonghao and Lin, Zi and Li, Zhuohan and Li, Dacheng and Xing, Eric and others},
  journal={Advances in neural information processing systems},
  volume={36},
  pages={46595--46623},
  year={2023}
}

@misc{yang2024causalbenchreview,
  title        = {A Critical Review of Causal Reasoning Benchmarks for Large Language Models},
  author       = {Yang, Linying and Shirvaikar, Vik and Clivio, Oscar and Falck, Fabian},
  year         = {2024},
  howpublished = {arXiv},
  eprint       = {2407.08029},
  archivePrefix= {arXiv},
  primaryClass = {cs.CL},
  url          = {https://arxiv.org/abs/2407.08029}
}

@inproceedings{causalbench2024,
  title={Causalbench: A comprehensive benchmark for evaluating causal reasoning capabilities of large language models},
  author={Wang, Zeyu},
  booktitle={Proceedings of the 10th SIGHAN Workshop on Chinese Language Processing (SIGHAN-10)},
  pages={143--151},
  year={2024}
}

@misc{openai2026comparemodels,
  author       = {{OpenAI}},
  title        = {Compare Models},
  year         = {2026},
  howpublished = {\url{https://platform.openai.com/docs/models/compare}},
  note         = {Last accessed February 2, 2026}
}
\bibliographystyle{iclr2026_conference}

\clearpage
\onecolumn
\appendix

\section{Appendix}

\subsection{Additional Experiments}

In this appendix, we provide robustness analyses to examine whether the semantic re-ranking behavior we observe is stable under a post-cutoff evaluation setting and across different LLM model variants.

\begin{table}[h]
  \centering
  \caption{Trading performance after training cutoff (entry date $>$ May 31, 2024).}
    \label{tab:training_cutoff}
  \begin{tabular}{lccc}
    \toprule
    \textbf{Metric} & \textbf{Statistical Approach} & \textbf{Hybrid Approach} & \textbf{Change} \\
    \midrule
    Win Rate & 61.8\% & 62.1\% & +0.3pp \\
    Avg Win & \$614 & \$656 & +6.8\% \\
    Avg Loss & -\$700 & -\$418 & \textbf{+40.3\%} \\
    \textbf{Total PnL} & \textbf{\$3,800} & \textbf{\$7,200} & \textbf{+89\%} \\
    \bottomrule
  \end{tabular}
\end{table}

\paragraph{Post-cutoff evaluation to mitigate lookahead bias.}
To mitigate potential lookahead from an LLM’s pretraining knowledge, we evaluate our method on a post-cutoff dataset consisting only of test periods after May~31,~2024, based on the publicly documented training data cutoff of the LLM \citep{openai2026comparemodels}. All other components of the pipeline are kept unchanged.

As shown in Table~\ref{tab:training_cutoff}, the hybrid approach continues to outperform the statistical baseline under post-cutoff evaluation. As in the main results, the average loss magnitude is reduced by 40.3\% (from \$700 to \$418), yielding an 89\% improvement in total PnL. These results reinforce the interpretation of the LLM as a semantic risk manager that systematically suppresses large downside losses by deprioritizing statistically fragile relationships.

\paragraph{Robustness across LLM model variants.} We further repeat the entire semantic re-ranking using an alternative model variant, \texttt{GPT-5-mini}, and observe the same loss-reduction behavior. This suggests that the risk-management effect is robust across model choices.

\subsection{Prompt Template for LLM-Based Semantic Filtering}
\label{app:prompt}

We use a fixed prompt format across all candidate pairs to ensure consistent semantic scoring and to avoid pair-specific tuning.

\begin{figure}[h]
  \centering
  \includegraphics[height=0.4\textwidth]{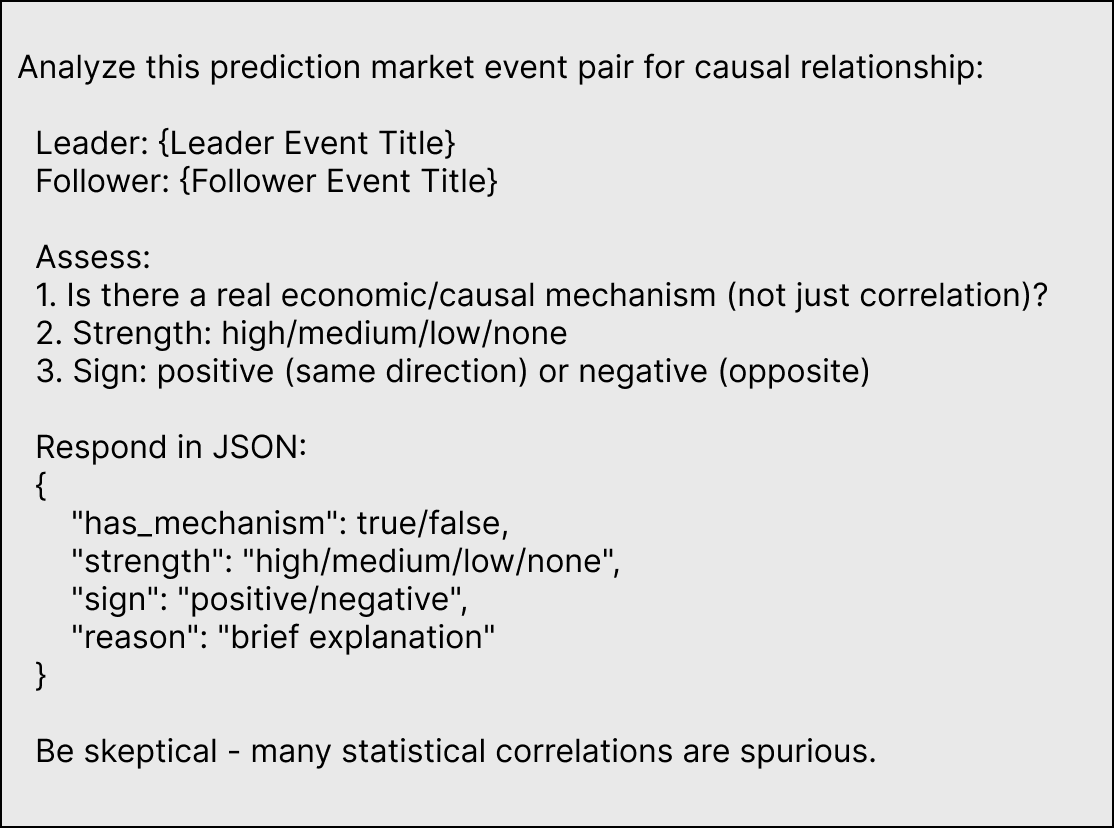}
  \caption{
  Prompt template used for LLM-based semantic filtering.
  Given a directed leader--follower event pair, the model assesses whether a plausible economic transmission mechanism exists (beyond correlation), assigns a strength level, and predicts the expected sign of co-movement, returning a structured JSON output.
  }
  \label{fig:prompt_template}
\end{figure}

\end{document}